\documentclass[twoside,twocolumn]{article}
\usepackage{multirow}
\usepackage[sc]{mathpazo} 
\usepackage[T1]{fontenc} 
\usepackage{microtype} 
\usepackage[english]{babel} 
\usepackage{graphicx}
\usepackage[hmarginratio=1:1,top=32mm,columnsep=20pt]{geometry} 
\usepackage[hang, small,labelfont=bf,up,textfont=it,up]{caption} 
\usepackage{booktabs} 
\usepackage{float}
\usepackage{lettrine} 
\usepackage{textcomp} 
\usepackage{enumitem} 
\setlist[itemize]{noitemsep} 

\usepackage{abstract} 

\usepackage{titlesec} 
\titleformat{\section}[block]{\Large\scshape\centering}{\thesection.}{1 em}{} 
\titleformat{\subsection}[block]{\large}{\thesubsection.}{1em}{} 
\titleformat{\subsubsection}[block]{\large\scshape}{\thesubsubsection.}{1em}{}[]

\usepackage{fancyhdr} 
\usepackage{titling} 

\usepackage{hyperref} 
\usepackage{verbatim}
\usepackage{amsmath}
\usepackage{nccmath}
\usepackage{multicol}

\pretitle{\begin{center}\Huge\bfseries} 
\posttitle{\end{center}} 
\title{\large Full Quaternion Representation of Color images: A Case Study on QSVD-based Color Image Compression} 
\author{%
\textsc{Alireza Parchami, Mojtaba Mahdavi} \\[1ex] 
\normalsize Department of Computer Engineering\\University of Isfahan, Iran \\ 
\normalsize {alirezaprm@mehr.ui.ac.ir, m.mahdavi@eng.ui.ac.ir} 
}

\date{} 
\renewcommand{\maketitlehookd}{%
\begin{abstract}

\noindent
For many years, channels of a color image have been processed individually, or the image has been converted to grayscale one with respect to color image processing. Pure quaternion representation of color images solves this issue as it allows images to be processed in a holistic space. Nevertheless, it brings additional costs due to the extra fourth dimension. In this paper, we propose an approach for representing color images with full quaternion numbers that enables us to process color images holistically without additional cost in time, space and computation. With taking auto- and cross-correlation of color channels into account, an autoencoder neural network is used to generate a global model for transforming a color image into a full quaternion matrix. To evaluate the model, we use UCID dataset, and the results indicate that the model has an acceptable performance on color images. Moreover, we propose a compression method based on the generated model and QSVD as a case study. The method is compared with the same compression method using pure quaternion representation and is assessed with UCID dataset. The results demonstrate that the compression method using the proposed full quaternion representation fares better than the other in terms of time, quality, and size of compressed files.
\\
\textbf{Keywords:}  Quaternion representation; Color image processing; Quaternion singular value decomposition; Color image correlation; Color image compression
\end{abstract}
}

\begin{document}

\maketitle


\section{Introduction}
\lettrine[nindent=0em,lines=3]{I} mage Processing and Computer Vision are widely used in many different applications and have a dominant role in industries and our lives. The increasing number of applications using color images makes image processing a crucial point in artificial intelligence. Therefore, we have witnessed rapid enhancement in different algorithms and applications derived from image processing and computer vision in many fields.
\par
Throughout history, colors in images have seen dramatic changes. During the 20th century, black-and-white images were mostly used. Since a black-and-white image only represents intensity levels, it forms a 2D matrix consists of real numbers. As a result, the processing procedure is not complicated due to the existence of one 2D matrix.
By color images, which each of them consists of three separate 2D matrices, the processing procedure would be controversial. Since there has been no holistic approach for color image processing that considers the image as one entity, two approaches are mostly used for color images that none of them is efficient enough. The first approach tries to convert a color image into a grayscale one and only cares about the intensity levels, which means it completely neglects the colors. The second approach, however, separates a color image into three scalar matrices to be processed individually by different algorithms and be masked by various filters. However, this approach is neither comprehensive nor efficient due to ignoring cross-correlation by channel separation.
\par
The idea of using quaternion numbers in color image processing was introduced in 1996 by Sangwine, using the discrete version of Ell\textquotesingle s Quaternion Fourier Transforms \cite{sangwine1996}. By this technique, image processing in a holistic way was made possible since quaternions empower us to consider an image as one matrix in a holistic space. This approach also enabled us to consider cross-correlation while processing color images. However, there were two main disadvantages that prevented this approach to be as much time- and space-efficient as it could be. Therefore, two other alternatives were introduced to alleviate the issues, but they also had some problems, which are completely discussed in Section 2: Related Works.
\par
The research reported in this paper aims to introduce a method for image processing in a holistic way using full quaternion numbers, which treats the color image as a vector field that has four highly correlated dimensions, in order to exploit the maximum potential of auto- and cross-correlation of color channels.
\par
In addition to introducing a model for full quaternion representation of color images, a compression method based on the proposed representation and Quaternion Singular Value Decomposition (QSVD) was presented in this paper as a case study. This method was evaluated by UCID dataset, and was compared with another compression method that uses pure quaternion numbers.
\section{Related Works}
\begin{figure}[h!]
\centering
  \includegraphics[width=0.5\textwidth]{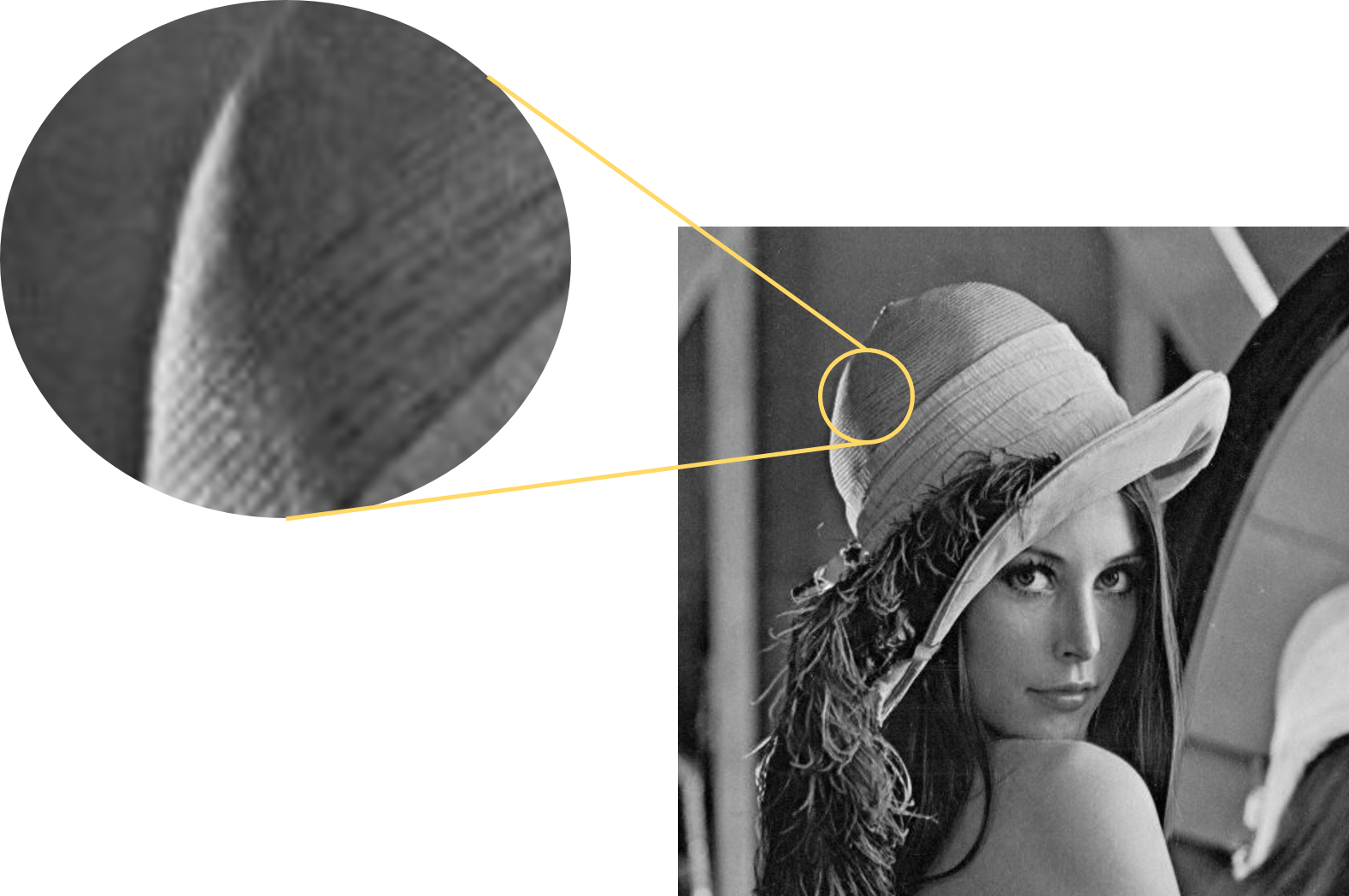}
  \caption{Auto-correlation of adjacent pixels in green channel}
  \label{fig:LenaCorrelation}
\end{figure}

It is believed that there are correlations between adjacent pixels of an image, whether color or grayscale one, and we can consider local correlations in different parts of an image. Chrominance and luminance change so slightly in adjacent pixels, and there is no dramatic difference between them. That is, each subpixel in R, G and B channels at coordination $(i, j)$ is correlated with its corresponding adjacent subpixels. This concept is also referred to as \emph{auto-correlation}, which indicates the similarity of a signal with itself, and it was defined for pure quaternion representation of color images \cite{Sangwine1999Hypercomplex}. Figure \ref{fig:LenaCorrelation} shows adjacent pixels of the green channel in a local area that are highly similar, and the difference from one subpixel to its neighbours is subtle.

\par
For color images, in addition to auto-correlation, there are correlations between the color channels. Capturing the inherent correlations between color channels and taking them into account, can indeed demonstrate more information than the conventional approach, which color channels were processed separately and the individual output results were combined. The concept of correlation between channels is referred to as \emph{cross-correlation}, which indicates the similarity of two distinct signals, and it was defined and proved applicable for quaternion representation of color images \cite{Sangwine1999Hypercomplex, Moxey2003Hypercomplex}. Using cross-correlation is one of the main motivations to adopt quaternions for digital color image processing.
\par
Many methods have tried to introduce new ways of processing to enhance the efficiency of image processing and computer vision concerning the correlation of color channels. When it came to color image processing, we mainly relied on grayscale images due to the higher inherent complexity of color images and the difficulty of considering the 3D space of color images as one entity. Therefore, two basic approaches were taken toward image processing and computer vision. In the first approach, a color image was transformed into a grayscale image by using either Rec. BT.601 \cite{Rec.BT.601}
\begin{equation*}
Gray = (0.299Red + 0.587Green + 0.114Blue)
\end{equation*}
 or Rec. BT. 709 recommended by ITU-R\footnote{International Telecommunication Union Radiocommunication Sector} \cite{Rec.BT.709}:
\begin{equation*}
Gray = (0.2126Red + 0.7152Green + 0.0722Blue)
\end{equation*}
By this conversion, the images was considered as a 2D grayscale matrix, and then the matrix was processed by desirable algorithms. In this case, none of the objects was distinguishable with their colors. Although this approach would be fast enough and easy to compute, it was not appropriate for many applications because the colors of objects can indeed indicate significant features, and the cross-correlation between three channels can offer useful information.
\par
The idea of using color channels for color image processing sparked by using separate channels for primary colors, which are red, green and blue. This approach, which tried to consider colors of objects, separated a color image into three distinct channels. Therefore, we had three 2D matrices as a consequence of having three channels for color images and treated each of them as a grayscale image. Whether we used neural networks or other image processing techniques, the cross-correlation between channels was not taken into consideration due to the channel separation in this approach. As a result, we lost data related to the cross-correlation of the color channels at the very beginning step of the process because we did not consider a color image as one unique entity.
\par
In 1998, Sangwine proposed a method based on quaternion numbers that applied the benefits of quaternion numbers into color image processing. In this method, a color image was considered as one matrix consists of pure quaternion numbers, and the matrix was processed in a holistic space. To convert an RGB image to quaternion matrix, RGB channels were simply put into the vector part of the quaternion matrix. Thus, a quaternion matrix $f_q(m,n)$, which represents a color image, was formed as\cite{sangwine1996,Pei2001}
\[ f_q(m,n) = f_R(m,n)i + f_G(m,n)j + f_B(m,n)k\]
where $f_R(m,n)$, $f_G(m,n)$ and $f_B(m,n)$ correspond to red, green and blue channels of the color image. By this representation, the scalar part of the matrix was actually considered as zero, which means a channel with no information. Thus, the final representation is like:
\begin{equation*}
\resizebox{0.94\hsize}{!}{$ f_q(m,n) = 0 + f_R(m,n)i + f_G(m,n)j + f_B(m,n)k$}
\end{equation*}
In the rest of this paper, we call this approach as \emph{pure quaternion representation} because it only deals with vector part of the quaternion matrix. As a consequence of the development of quaternion mathematical operations, many applications such as splicing detection \cite{Li2017Image}, watermarking \cite{Li2018Color, Chen2018Quaternion, Hosny2018Robust, Bas2003Color, Tsougenis2014Adaptive, Wang2013A}, filter implementation \cite{Wang2014A,  Yasmin2016Quaternion,Sangwine2000Colour}, image classification \cite{Zeng2016Color}, image sparse representation \cite{Xu2015Vector} and quaternion-type moment \cite{Chen2015Color} have treated color images holistically as a vector field using the pure quaternion representation for the last 20 years.
\par
This representation empowers us to use various quaternion mathematical operations for processing the image matrix comprehensively concerning the auto- and cross-correlation. However, it does not benefit from all components of a quaternion matrix and only concerns about the vector part, which leads to the fact that the maximum potential of quaternions is not being used. To illustrate more, with image processing using pure quaternion numbers, we have to deal with a new channel filled with zeros for the real component of the quaternion matrix. This way, a three-channel color image is actually considered as part of a four-channel image with an extra black channel that has no correlation with others and is completely unrelated with each of the channels conveying specific information. It can make procedures of processing more complicated since quaternion mathematical operations, such as QDCT or QSVD, consider all components of the quaternions and have to take a new channel into consideration which is irrelevant to others. Furthermore, the method increases both the size of the matrix and time of processes due to adding a new channel with the same size of the other channels \cite{Assefa2011The}, which leads to having four 2-D matrices with the same number of columns and rows as the original image.
\par
As an alternative, a new type of representation was introduced based on trinion numbers, which have one real and two imaginary units. In this approach, a color image was represented as \cite{Assefa2011The}:
\begin{equation*}
\resizebox{0.94\hsize}{!}{$ f_q(m,n) = f_R(m,n) + f_G(m,n)i + f_B(m,n)j$}
\end{equation*}
This representation aimed to avoid adding a new channel and assign one component to each of the RGB channels. Some mathematical operations were also generalized to trinions for this purpose. However, only few color image processing works adopted trinion representation because the theory of trinions has not enough developed compared with the theory of quaternions \cite{Assefa2011The}.
\par
As another alternative, some scientists developed another method that tries to use all four components of quaternion numbers. In this method, they captured more information and measured the depth of color images and used quaternion numbers to combine both color and depth information for different purposes such as RGB-D object recognition \cite{Chen2016Quaternion} or color image splicing detection \cite{Chen2017Quaternion}. Therefore, the quaternion matrix was formed as:
\begin{equation*}
\begin{split}
 f_q(m,n) &= f_D(m,n) + f_R(m,n)i  \\
 &+ f_G(m,n)j + f_B(m,n)k
\end{split}
\end{equation*}
where $f_D(m,n)$ indicates the depth information. Although it has been proved to be a good technique for adopting all components of quaternion numbers, it needs additional devices to measure the depth information while capturing the photos such as Kinect, and it is not possible to use this method with ordinary cameras \cite{Chen2016Quaternion, Chen2017Quaternion}. This is the main reason that the full quaternion representation with RGB-D information has not been used widely for image processing.
\par
With regard to the application of QSVD, color image compression is one of the fields that quaternions can be used, and singular value decomposition (SVD) is a classical mathematical tool that has been used for a long time to compress grayscale and color images. Quaternion singular value decomposition (QSVD) is the quaternion version of this useful tool and has exploited in color image processing \cite{sangwine1996, Ell2007Hypercomplex}. Therefore, color image compression using QSVD became possible, and it has been proved to be an appropriate application. Both SVD-based image compression using three separated channels and QSVD-based image compression using pure quaternion representation were implemented and compared with each other by Ying Li \cite{Li2017Comparison}. Although the comparison was entirely on the favour of SVD-based color image compression, the main downsides of QSVD-based color image compression were proved to be the long time required for QSVD calculation and lower compression rate (CR). In this paper, QSVD is used to compress color images using the full quaternion representation to put our method into practice. To solidify our compression method with full quaternion numbers, we implemented this compression method on both pure quaternion and full quaternion representations to compare them with each other.
\section{Methodology}
As mentioned in section 2, the pure quaternion representation processed color images comprehensively concerning the auto- and cross-correlation. However, the extra fourth dimension not only diluted the correlation of color channels but also increased processing time and matrix size. The trinion numbers tried to use correlation more effectively than the previous work, but its mathematical concepts have not been well-developed. Moreover, using RGB-D color space for quaternion representaiton required special devices that are not feasible in most of the times.
\par
Regarding using all components of quaternion numbers, we sought for an approach which converts an RGB image into a full quaternion matrix to use quaternion mathematical operations and process images comprehensively. By this method, not only do we take full advantage of quaternion mathematical concepts and treat with color image as a full quaternion matrix, but also we use four correlated 2D matrices with a reduction in the number of columns to form the quaternion matrix that culminates in a cutback in the size and time of processing. Besides, no additional devices is needed in this approach to transform a color image into a full quaternion matrix. In the rest of this paper, we call this approach as \emph{full quaternion representation}.
\par
In the rest of this section, at first, a broad overview of the key mathematical concepts and tools used in this research are provided. Afterwards, we propose a method using an autoencoder to generate a global model and then, use the model to transform an RGB image into a quaternion one. subsequently, quaternion image compression is introduced as a case study of the proposed method.
\subsection{Background}
\subsubsection{Quaternion Numbers}
Quaternion numbers, also known as hyper-complex numbers, consist of one scalar and one vector part that has three imaginary units. In fact, a quaternion $q$ is an element of the 4-D normed algebra with basis ${1, i, j, k}$. Quaternion number $q$ is represented in the form
\[ q = a + bi + cj + dk \]
where $a, b, c, d \in \mathbb{R}$ are called components and $i, j, k$ are fundamental quaternion units \cite{Bihan2003Quaternion}. Rules of the product of fundamental units are defined by

\begin{equation}
\begin{aligned}
ij &= -ji = k \\
jk &= -kj = i \\
ki &= -ik = j \\
ijk &= i^2 = j^2 = k^2 = -1
\end{aligned}
 \label{eqn:QuaternionUnitRelations}
\end{equation}

A quaternion $q \in \mathbb{H}$ can\footnote{Classically denoted by $\mathbb{H}$ in honor of Sir W.R. Hamilton who discovered them in 1843} be decomposed into a scalar part $S(q)$ and a vector part $V(q)$:
\[ q = S(q) + V(q)\]
where
\[ S(q) = a, \ V(q) = bi + cj + dk\]
The quaternion $q$ is called \emph{pure quaternion} if $S(q) = 0$. It is normally known as a real number if $V(q) = 0$, and it is called \emph{full quaternion} if none of the scalar and vector parts  are zero\footnote{$V(q) \neq 0$ if at least one of the components $b, c, d$ is non zero.}. Note that the multiplication of two quaternion numbers $p, q \in \mathbb{H}$ is not commutative due to the relation between fundamental units in (\ref{eqn:QuaternionUnitRelations}), but it is associative for three quaternion $p, q, r \in \mathbb{H}$ \cite{Ell2014Book}:
\[ p \times q \neq q \times p\]
\[(p \times q) \times r = p \times (q \times r)\]
Moreover, the conjugate of a quaternion, which is denoted by $\bar{q}$ or $q^*$, is defined as
\[\bar{q} = a - bi - cj - dk \]
\subsubsection{Autoencoder}
\begin{figure}[!h]
\centering
  \includegraphics[width=0.45\textwidth]{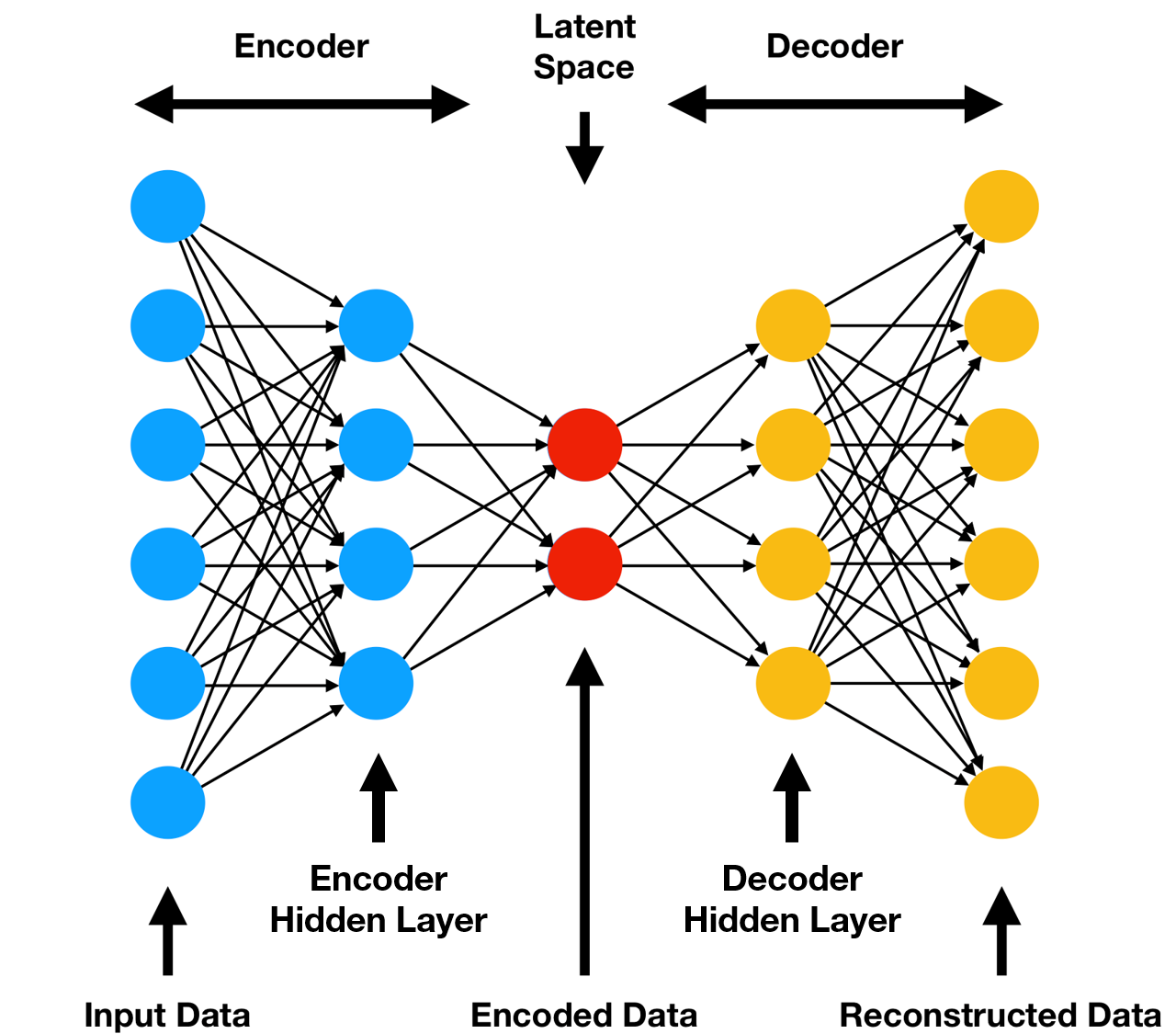}
  \caption{Feedforward autoencoder with 2 nodes in latent space}
  \label{fig:AutoencoderFigure}
\end{figure}

A Feedforward autoencoder is a type of artificial neural networks. It has the same number of nodes in input and output, and it tries to encode data in the latent layer with the purpose of reconstructing the input based on the encoded data. The connections do not form a cycle, and the nodes of the middle layers would vary widely. It is an unsupervised learning process which pursues encoding inputs into a desirable number of nodes efficiently to recover them as accurately as possible. Figure \ref{fig:AutoencoderFigure} shows the elements of an autoencoder.

\par
In an autoencoder, the existence of hidden layers is not compulsory. That is, the number of encoder and decoder hidden layers would vary widely according to researchers' purpose. However, the latent space is the key layer since input data are encoded in this layer and then, the encoded data are used for processing.
\subsubsection{Quaternion Singular Value Decomposition (QSVD)}
Singular Value Decomposition (SVD) is a well-known linear algebra technique which represents a matrix $A$ in a coordinate system where the covariance matrix is diagonal. SVD has been proved as an image compression method, which generalizes the eigendecomposition of a matrix by the extension of polar decomposition. The SVD of a matrix $A \in \mathbb{R} ^{m \times n}$ is given as:
\[ A = U  \Sigma V^T \]
where $U\in\mathbb{R} ^{m \times m}$ is left singular vector, $\Sigma \in\mathbb{R} ^{m \times n}$ stands for singular values, and $V\in\mathbb{R} ^{n \times n}$ is right singular vector. This concept has also been extended to quaternion numbers, known as quaternion singular value decomposition (QSVD) \cite{Pei2003QSVD}. The proof of Quaternion Polar Decomposition and Quaternion Singular Value Decomposition is given in \cite{Zhang1997} .
\par
There is a QSVD for every quaternion matrix $Q \in \mathbb{H} ^{N \times M}$ given as:
\[ Q = U  \Sigma V^\triangleleft \]
where $^\triangleleft$ stands for conjugate transposition, $U\in\mathbb{H} ^{N \times N}$ indicates left singular vector and $V\in\mathbb{H} ^{M \times M}$ indicates right singular vector. $U$ and $V$ are unitary quaternion matrices, which means
\[S(UU^\triangleleft) = S(VV^\triangleleft) = I\]
and
\[V(UU^\triangleleft) = V(VV^\triangleleft) = O\]
where $S(x)$ and $V(x)$ denote the scalar and vector parts of the quaternions, respectively. In the notation,  $I$ is the identity matrix (diagonal matrix of ones) and $O$ is a matrix contains only zeros \cite{Zhang1997, Bihan2003Singular}. The same as SVD, $\Sigma \in\mathbb{R} ^{m \times n}$ indicates singular values of the quaternion matrix $Q$.
\par
Having a compressed image based on SVD, the number of preserved columns of $U, \Sigma$ and $V$ indicates the level of compression. That is, we can modify the quality of the compressed image and the size of the compressed file by matrix truncation, which means adopting the $t \in \mathbb{N}$ largest singular values and the corresponding vectors of the left and right unitary matrices, and leaving the rest of the values as zero \cite{Li2017Comparison}.  As the real values of $\Sigma$ arranged in decreasing magnitude order along the diagonal, the first $t$ values carry the most significant decomposition values and have a vital role in image reassembling.
\subsection{Proposed Method}
Taking local correlation into account, we looked for a relationship that enables us to represent a color image with four 2D real matrices, which all of them are highly correlated, to create a full quaternion matrix.  For this purpose, we considered a pair of adjacent pixels at coordinations $(i,j)$ and $(i,j+1)$ where $i$ and $j$ indicate the position of row and column of pixels, respectively. Because of colorfulness of the image, we have 3 sub-pixel values in RGB color space for each pixel. Thus, there are six values for the pair above:
\begin{equation*}
\begin{aligned}
R(i,j) &, G(i,j) , B(i,j) \\
R(i,j+1) &, G(i,j+1) , B(i,j+1)
\end{aligned}
\end{equation*}
\par
\begin{figure}[!h]
\centering
  \includegraphics[width=0.5\textwidth]{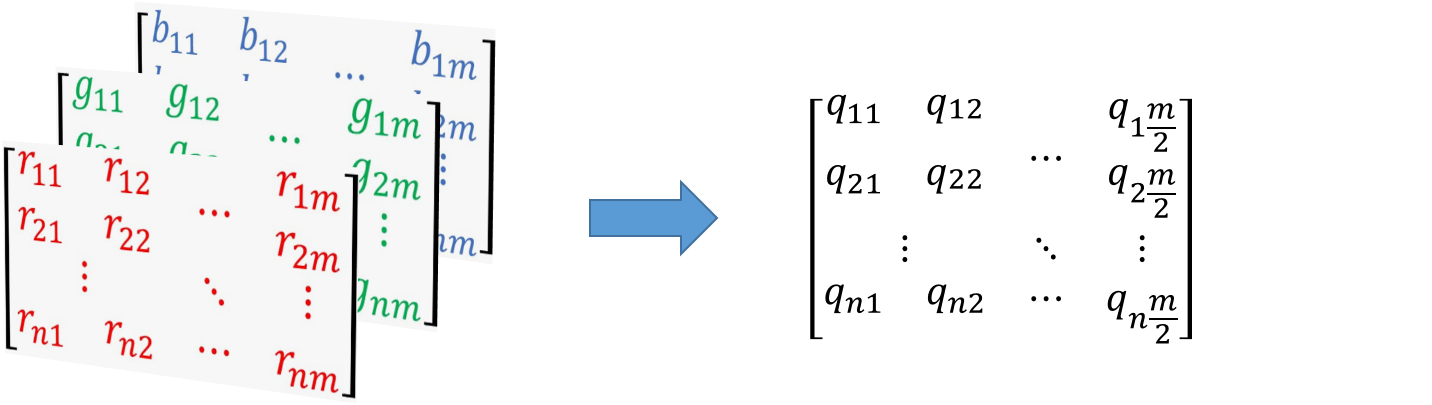}
  \caption{Dimensions of RGB channels and full quaternion matrix}
  \label{fig:RGBtoQ}
\end{figure}

We used a full connect feedforward autoencoder to find coefficients and biases for these six subpixels to generate a model that transforms each pair of pixels into four numbers that can form a full quaternion number. Using this model for all pairs of adjacent pixels in a color image, we formed a quaternion matrix with the same number of rows as the original image and the half number of columns because each pair of adjacent pixels in RGB color space at coordinations $(i,j)$ and $(i,j+1)$ is transformed into one quaternion number. Figure \ref{fig:RGBtoQ} shows the size of both original and quaternion matrices with differences in their second and third dimensions. Original matrix is considered as a 3-D matrix of integer numbers with the size of $N \times M$, while the quaternion matrix is a 2-D matrix of quaternion numbers with the size of $N \times \frac{M}{2}$. 
It is notable that prior to transforming a color image into a full quaternion matrix, if the image has odd number of columns, replicate padding\footnote{In \emph{replicate padding},values outside the boundary are set equal to the nearest image border value\cite{Gonzalez2018Digital}.} is used to make the number of columns even. This is because we have to consider pixels two by two in order to form the quaternion matrix.
\par
\begin{figure}[hb!]
\centering
  \includegraphics[width=0.5\textwidth]{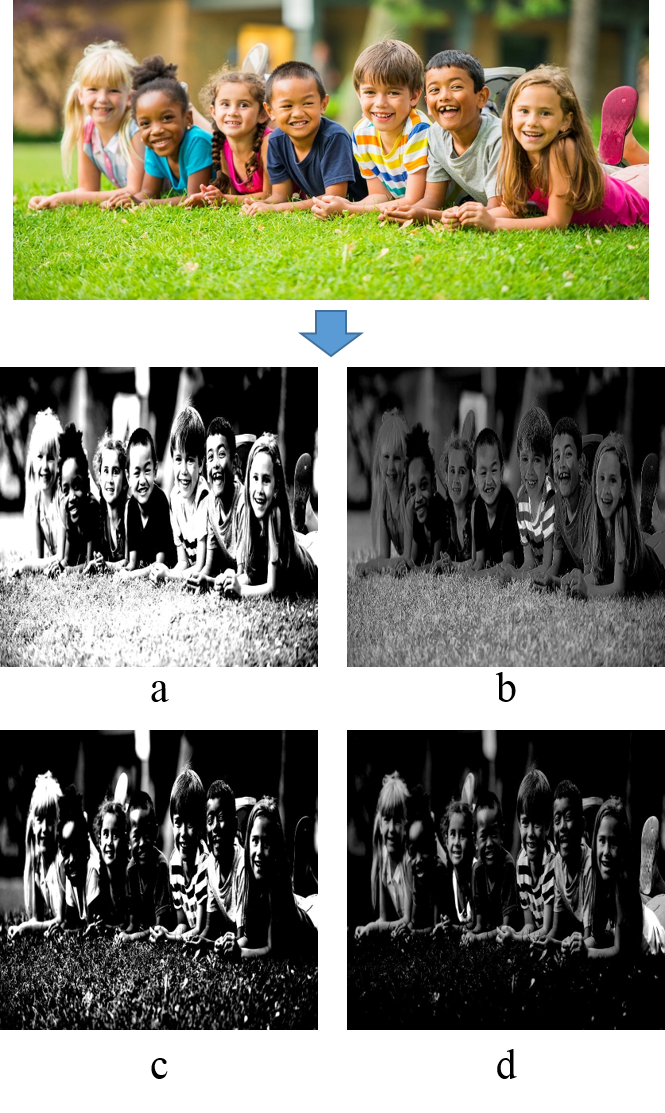}
  \caption{Quaternion components of a color image}
  \label{fig:planes}
\end{figure}

Like every color space that each component carries particular features, each component of the quaternion matrix contains specific information from the original image. Figure \ref{fig:planes} shows each component of the quaternion matrix that was acquired from the generated model. After processing the quaternion matrix and implementing algorithms with quaternion mathematical operations, the same model was used to revert the full quaternion matrix to RGB color space in order to use the image in normal applications. That is, we decoded the encoded data to reach the RGB pixels. 

\par
\begin{figure}[h]
\centering
  \includegraphics[width=0.5\textwidth]{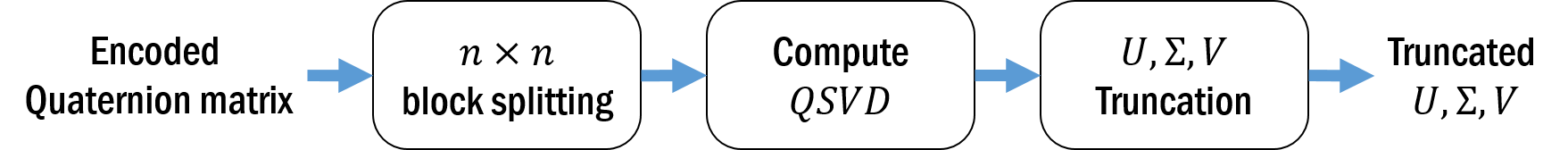}
  \caption{Quaternion Matrix Compression Steps}
  \label{fig:QCompressionMethods}
\end{figure}
\begin{figure}[h!]
\centering
  \includegraphics[width=0.5\textwidth]{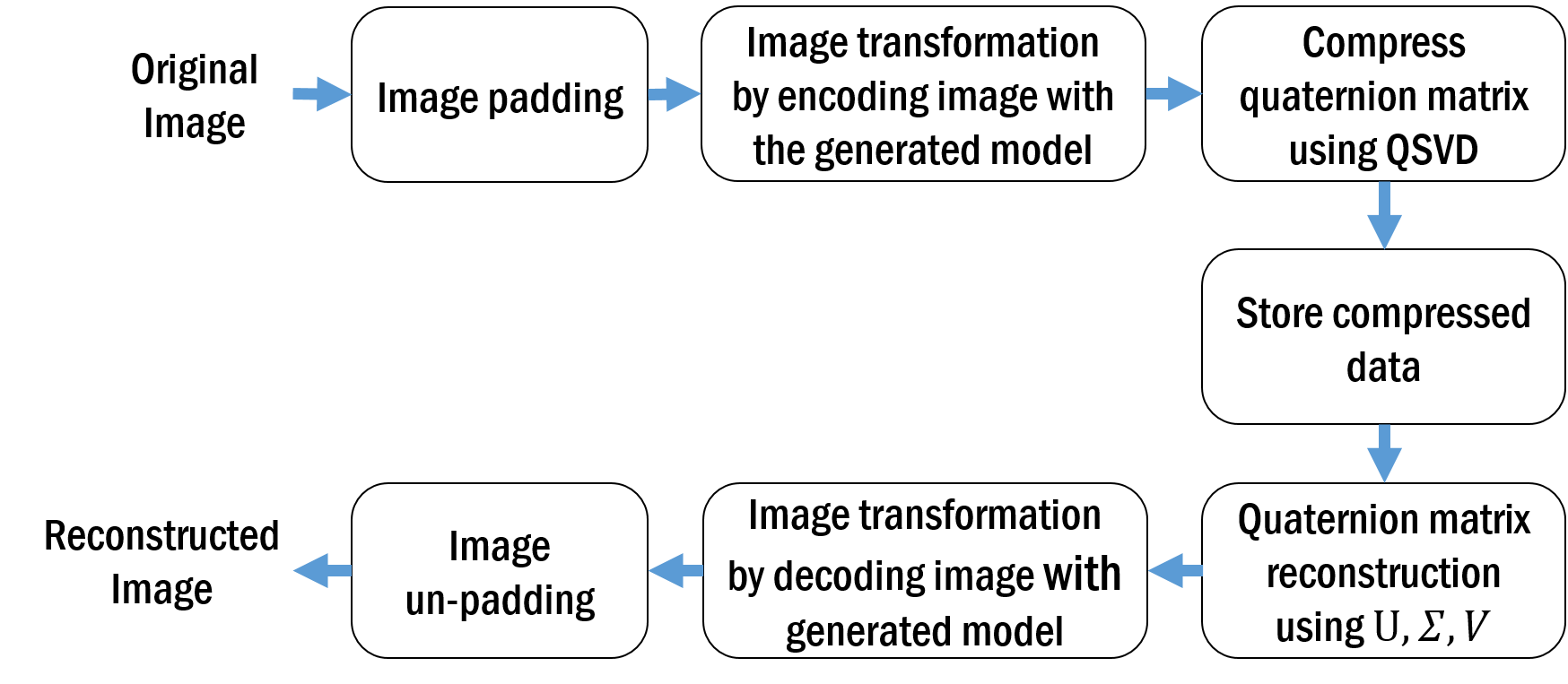}
  \caption{Processing Steps of Encoding and Decoding an RGB Image}
  \label{fig:QEncodingDecodingProcess}
\end{figure}

As a case study, image compression can be used as an appropriate practical application for representing an RGB image with a quaternion matrix \cite{Bihan2003Quaternion, Luo2010Color, Mahdu2018Image}. Figure \ref{fig:QCompressionMethods} shows the diagram of color image compression quaternion numbers.
Compression was done in three steps. As the calculation time of both of the SVD and QSVD would increase exponentially with higher resolution of images, block splitting was done to enhance the calculation time. Therefore, we tried to reach the most suitable block size concerning the QSVD calculation time and the quality of image. QSVD was calculated for each block and it was truncated by $t$, which is our modifiable measure for compression. Finally, the truncated $U$, $\Sigma$ and $V$ matrices were compressed by RAR algorithm using WinRAR software by setting compression method as best. Figure \ref{fig:QEncodingDecodingProcess} illustrates on the steps of QSVD-based color image compression utilizing the generated model and the proposed full quaternion representation.

The quaternion matrix can be reconstructed by the formula of:
\[Reconstructed Quaternion Matrix= U  \Sigma V^\triangleleft\]
Finally, the full quaternion matrix was decoded into a 3-D matrix of RGB by using the model generated by the designed autoencoder.
\subsection{Methodology Parameters}
We used a full connect feedforward autoencoder to find a global model for color image conversion into a full quaternion matrix. The autoencoder had 6 nodes for input and output layers as it received and predicted 6 subpixels of two adjacent color pixels as mentioned. It had 4 nodes in latent space and there was no hidden layer in the encoder and decoder. We also used \emph{pure linear} activation function and \emph{L2 regularization} method  to train autoencoder with UCID dataset, so the model could be known as a regression model, which can also be considered as a lossy convertor. 60\% of the available UCID dataset was used for training and the rest of it was used for testing\footnote{The official website of UCID dataset has been out of access, and there is no other official alternative source for this dataset. The only available website is \url{http://jasoncantarella.com/downloads}. However, the dataset uploaded on this website is incomplete and contains only the first 886 images of the dataset. Therefore, we have succeeded to train the autoencoder with 536 randomly selected images (60\% of the available dataset) and test the proposed method with the remaining 350 images of UCID dataset. } \cite{Schaefer2003UCID}. 
\section{Results}
The quality of a reconstructed image is the primary key to evaluate the proposed model and compression method. We used 40\% of the available UCID dataset for assessing the model and compression method since UCID dataset provides a wide range of raw images in many different categories. We also used three methods for measuring the performance of the model for color iamge transformation to a quaternion matrix, namely SSIM, PSNR and MSE. Moreover, to evaluate the quaternion color image compression, we used PSNR, Compression Ratio (CR) and MSE for each of the color channels. The methods for measurement are the following four.

\begin{description}
\item [ $\bullet$ MSE] calculated for each channel separately and is given as:
\end{description}
\begin{equation*}
MSE = \frac{1}{N * M} \sum_{i=0}^{N-1} \sum_{j=0}^{M-1} {[I_1(i,j) - I_2(i,j)]}^2
\end{equation*}
Where $N$ and $M$ are the number of rows and columns of matrix , and $I1$ and $I2$ are the input and output matrix.

\begin{description}
\item [ $\bullet$ PSNR] which is given as:
\end{description}
\begin{multline*}
PSNR = 10\log_{10}{\frac{MAX_I^2}{MSE}} \\
= 20\log_{10}{MAX_I} - 10\log_{10}{MSE}
\end{multline*}
Where $MAX_{I}$ is the maximum possible value in matrix I, which is 255 for 24-bit color (true color) images.

\begin{description}
\item [ $\bullet$ SSIM] which is given as:
\end{description}
\begin{equation*}
SSIM = \frac{(2\mu_{I_1}\mu_{I_2} + C_1)(2\sigma_{{I_1}{I_2}} + C_2)}{(\mu_{I_1}^2 + \mu_{I_2}^2 + C_1) (\sigma_{I_1}^2 + \sigma_{I_2}^2 + C_2)}
\end{equation*}
Where $I_1$ and $I_2$ are original and reconstructed images; $\mu_{I_1}$ and $\mu_{I_2}$ are the average of $I_1$ and $I_2$ respectively; $\sigma_{I_1}^2$ and $\sigma_{I_2}^2$ are the variance of $I_1$ and $I_2$ respectively; $\sigma_{{I_1} {I_2}}$ is the covariance of $I_1$ and $I_2$; and
\begin{equation*}
C1=(K_1L)^2, \;\; C2=(K_2L)^2
\end{equation*}
with $K_1=0.01$, $K_2=0.03$ and $L = 255$.

\begin{description}
\item [ $\bullet$ Compression Ratio] which is straightforwardly given as:
\end{description}
\begin{equation*}
CR = \frac{Uncompressed \ Size}{Compressed \ Size}
\end{equation*}
\subsection{Transformation}
Using a global model for encoding 2 adjacent pixels into one quaternion number based on local correlation leads to a lossy conversion. Although an adequate model and high correlated pixels can compensate for the reduction of the quality of reconstructed images, the act of encoding and decoding are lossy. Table \ref{tab:ModelAssessment} shows the results of PSNR, SSIM and MSE of the generated model performed on the UCID dataset. Images were straightly encoded and decoded without any additional filters or processes.

\begin{table}[!h]
\begin{tabular}{|l|c|l|l|l|}
\hline
\multirow{2}{*}{PSNR} & \multirow{2}{*}{SSIM}       & \multicolumn{3}{c|}{MSE} \\ \cline{3-5} 
                      &                             & Red    & Green  & Blue   \\ \hline
41.2496               & \multicolumn{1}{l|}{0.9931} & 1.4578 & 0.7034 & 1.3184 \\ \hline
\end{tabular}
\caption{Assessment of the generated model on UCID dataset}
\label{tab:ModelAssessment}
\end{table}

The results in table \ref{tab:ModelAssessment} are good indicative of this fact that the act of encoding and decoding does not have distinguishable impact on images and can not be noticed by humans due to its low loss. The small values of MSE, especially in green channel, demonstrate that the differences between original and reconstructed pixels are pretty low, which can even be considered as a suitable method for image processing by machines and computers.
\par
\begin{figure}[h!]
\centering
  \includegraphics[width=0.5\textwidth]{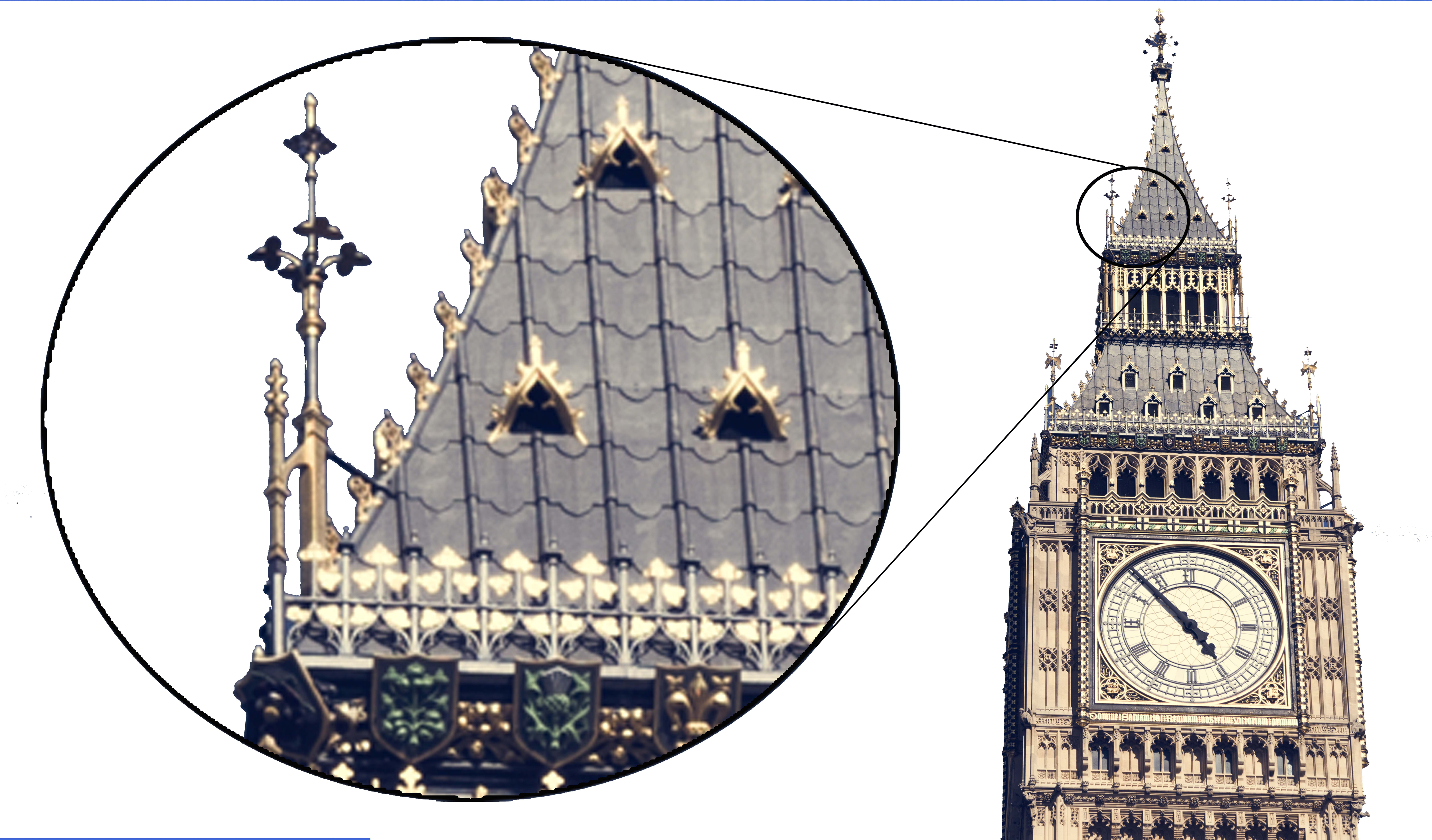}
  \caption{local correlation in edges}
  \label{fig:BinBenClock}
\end{figure}

Since the model is actually a type of a regression model, edges of images needs closer attention. We considered Big Ben in figure \ref{fig:BinBenClock} for model performance testing on photos full of edges. The sheer number of edges in this image makes it a perfect choice for this test. PSNR of 43.5667 and SSIM of 0.9990 show that the model encoded the photo into a full quaternion matrix and reconstructed the RGB image perfectly. The reason for this low amount of loss stands for the fact that the state of colors does not change dramatically in edge areas. That is, the chrominance and luminance change marginally in several arrays of pixels, and the auto-correlation and cross-correlation in the edges are still high. Therefore, the model performed well in these situations, and we can still count on the high correlation of corresponding subpixels in adjacent pixels.

\subsection{Quaternion Image Compression}
In this section,  PSNR, MSE and Compression Ratio are presented for the proposed full quaternion compression method. The same measures for pure quaternion compression method are taken in section V: Discussion, where both methods are compared with each other.
\par
\begin{figure}[h!]
  \includegraphics[width=0.5\textwidth]{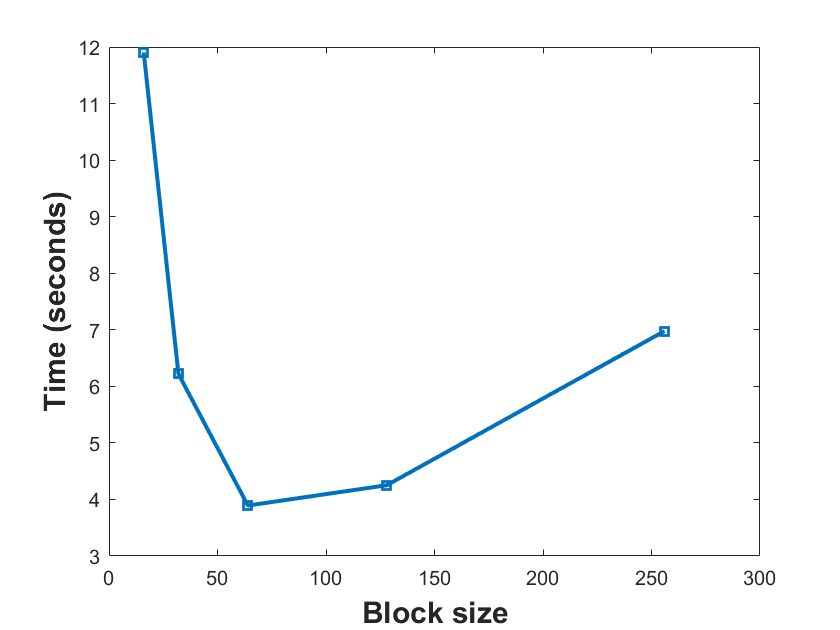}
  \caption{QSVD Calculation time}
  \label{fig:averageQSVDtime}
\end{figure}

In the proposed compression method, the calculation time for QSVD would be a discussing area. As QSVD considers all components of quaternion numbers comprehensively, the time of QSVD would last more than the calculation time of SVD on three separated 2D - matrices that contain only real numbers. Figure \ref{fig:averageQSVDtime} shows the calculation time of QSVD of the proposed full quaternion representation splitted into $n \times n $ blocks, with $n=16,32,64,128,256$. According to this figure, the most time-efficient block size for QSVD calculation is 64, which needs 3.8924 seconds on average using Intel\textregistered \space Core\textsuperscript{TM} i5-7400 processor, Matlab\textregistered \space R2017b, and Quaternion toolbox for Matlab\cite{Sangwine2020Quaternion}.

\par
\begin{figure}[h!]
  \includegraphics[width=0.5\textwidth]{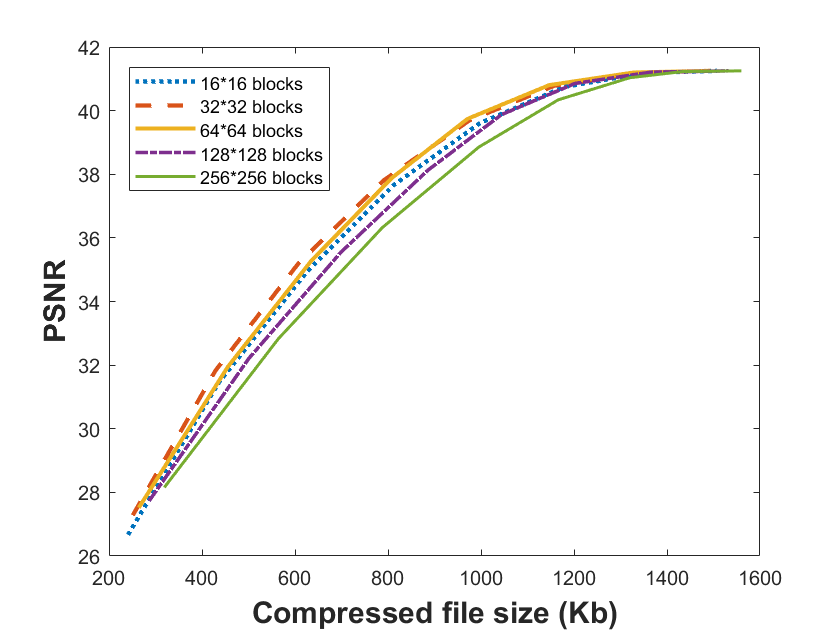}
  \caption{PSNR and compressed size of full quaternion compression (Higher is better in the same compressed file size)}
  \label{fig:FullQuaternionPSNRCompression}
\end{figure}

Figure \ref{fig:FullQuaternionPSNRCompression} depicts the results of measuring PSNR for compressed images. The size of compressed images were modified by adopting different $t$ for truncation of $U, \Sigma$ and $V$. Greater $t$ equals more preserved data in the matrices $U, \Sigma$ and $V$, which leads to higher quality and bigger file size. Based on this chart, block sizes of $32 \times 32$ and $64 \times 64$ have the best results because they have higher PSNR in the same compressed file size.

\par
\begin{figure}[h!]
  \includegraphics[width=0.5\textwidth]{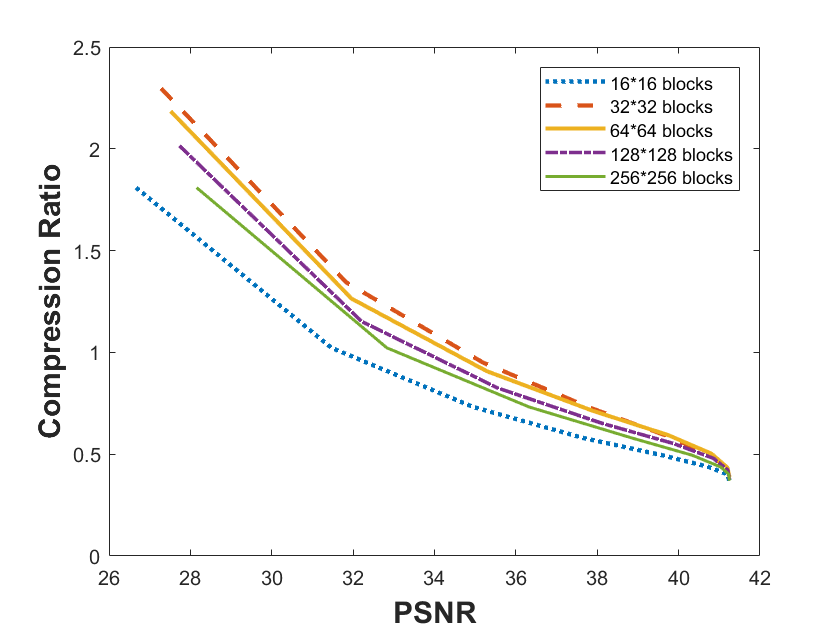}
  \caption{Compression ratio of full quaternion compression (Higher is better in the same PSNR)}
  \label{fig:FullQuaternionCompressionRatio}
\end{figure}

Compression ratio (CR) depends on $n$, which indicates the size of blocks for block splitting, and $t$, which indicates the number of preserved columns after matrix truncation. Figure \ref{fig:FullQuaternionCompressionRatio} demonstrates the CR measure for various $t$\textquotesingle s and $n$\textquotesingle s.

\par
\begin{figure}[h!]
  \includegraphics[width=0.5\textwidth]{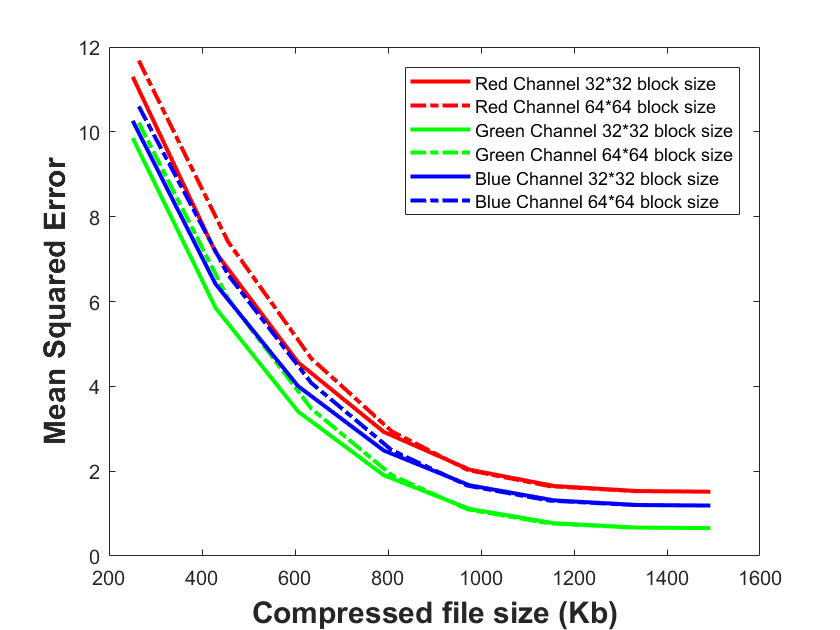}
  \caption{MSE of color channels for $32 \times 32$ and $64 \times 64$ block sizes (Lower is better in the same file size)}
  \label{fig:FullQuaternionMSE}
\end{figure}

Taking all measures above into account, the $64 \times 64$ and $32 \times 32$ block sizes seem to be the most efficient choices because they have faster QSVD calculation time and higher PSNR in the same file size. The compression method using $64 \times 64$ blocks is faster while the method using $ 32 \times 32$ blocks has a higher quality in the same compressed file size for PSNR lower than 38. As a result of better performance of these two block sizes, MSE for red, green and blue channels are given only for $32 \times 32$ and $64 \times 64$ block sizes in the figure \ref{fig:FullQuaternionMSE}.

\par
\begin{figure}[h!]
  \includegraphics[width=0.5\textwidth]{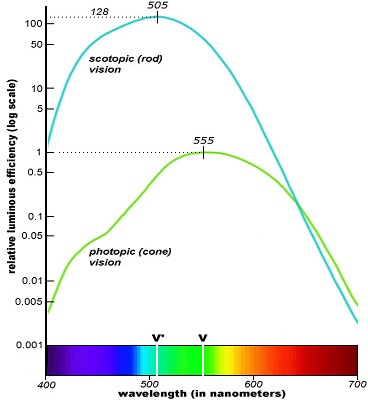}
  \caption{Photopic and Scoptopic sensitivity functions \cite{Kaiser1996Human}}
  \label{fig:EyeSensitivity}
\end{figure}

Since the Cone cells are responsible for color perception in daylight (Photopic), human eyes are more sensitive to greenish-yellow light than any other color because it stimulates medium and long cone types. However, in weak light, Rod cells are more brilliant for color perception (Scoptopic), and they are more sensitive to blueish-green wavelength. These two facts, which are shown in figure \ref{fig:EyeSensitivity}, contribute to the notion that human eyes are more sensitive to green color than blue and red \cite{handprint}. Based on the table \ref{tab:ModelAssessment} and figure \ref{fig:FullQuaternionMSE}, green, which is the most significant color for human eyes, has the least amount of error than the other colors in our proposed full quaternion representation and quaternion compression method, and this makes both of the proposed representation and compression very useful for common photography.
\section{Discussion}
In this section, the pure quaternion and full quaternion compression methods are compared with each other. The same as previous section, the QSVD calculation time, PSNR, compression ratio (CR) and MSE are evaluated.
\par
\begin{figure}[h!]
  \includegraphics[width=0.5\textwidth]{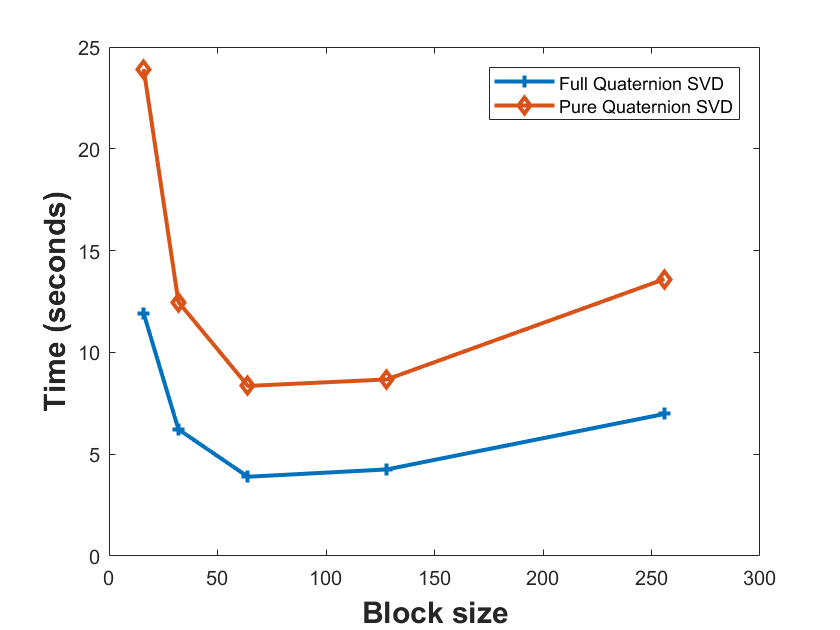}
  \caption{QSVD time for Full Quaternion and Pure Quaternion compression methods}
  \label{fig:comparisonQSVDtime}
\end{figure}

According to figure \ref{fig:averageQSVDtime}, the QSVD calculation would take considerable time. However, since our full quaternion method reduces the number of columns of images by half, it needs much less time for QSVD calculation compared with pure quaternion approach. As mentioned in section II, the act of adding a new channel with the same size as original channels in pure quaternion representation increases the time of processing. The figure \ref{fig:comparisonQSVDtime} indicates the comparison of QSVD calculation time for both of the discussed methods considering block sizes of 16, 32, 64, 128 and 256.

Since the QSVD calculation time is unreasonably high for block sizes of $16 \times 16$ and $256 \times 256$, and results of $64 \times 64$ blocks are better and faster than $128 \times 128$ blocks, we continued our tests for block sizes of $32 \times 32$ and $64 \times 64$.
\par
\begin{figure}
  \includegraphics[width=0.5\textwidth]{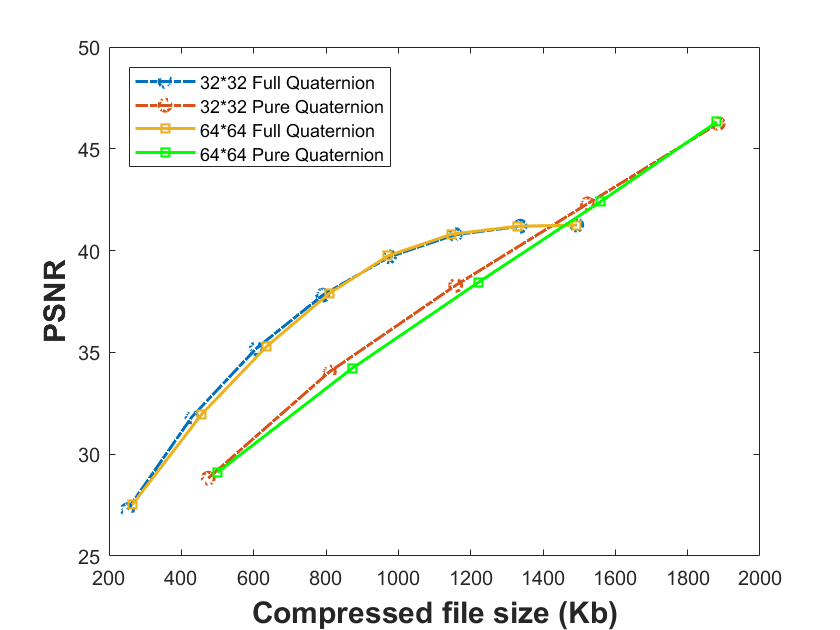}
  \caption{Quality and size of both of the compression methods}
  \label{fig:comparisonFileSize}
\end{figure}
\begin{figure}
  \includegraphics[width=0.5\textwidth]{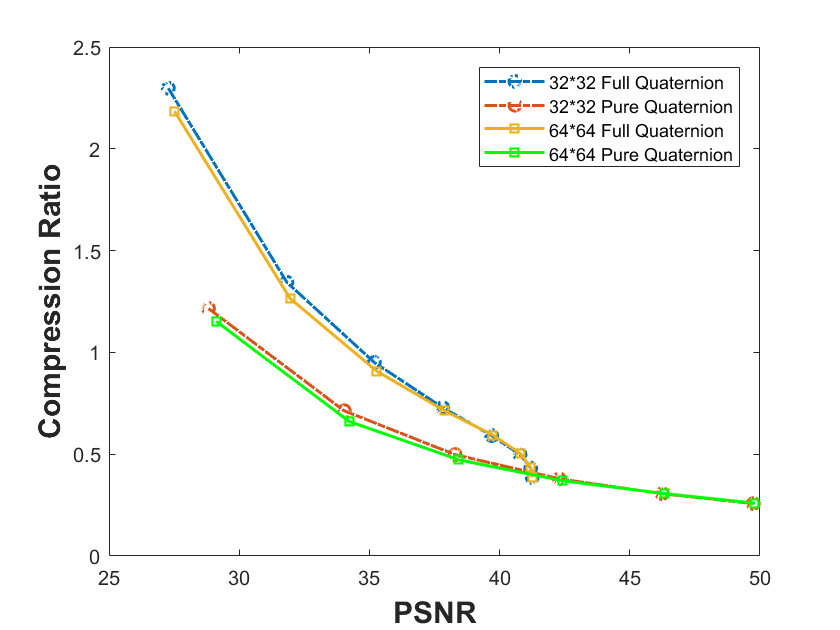}
  \caption{CR of full quaternion and pure quaternion compression methods}
  \label{fig:comparisonCR}
\end{figure}

The compressed file sizes related to PSNR for both compression methods are given in figure \ref{fig:comparisonFileSize}. Moreover, figure \ref{fig:comparisonCR} demonstrates the CR for both pure quaternion and full quaternion compression. Based on these charts, it is apparent that in the same PSNR, full quaternion compression method fares better than pure quaternion one because the data are compressed more in the former method.

\par
\begin{figure}[h]
  \includegraphics[width=0.5\textwidth]{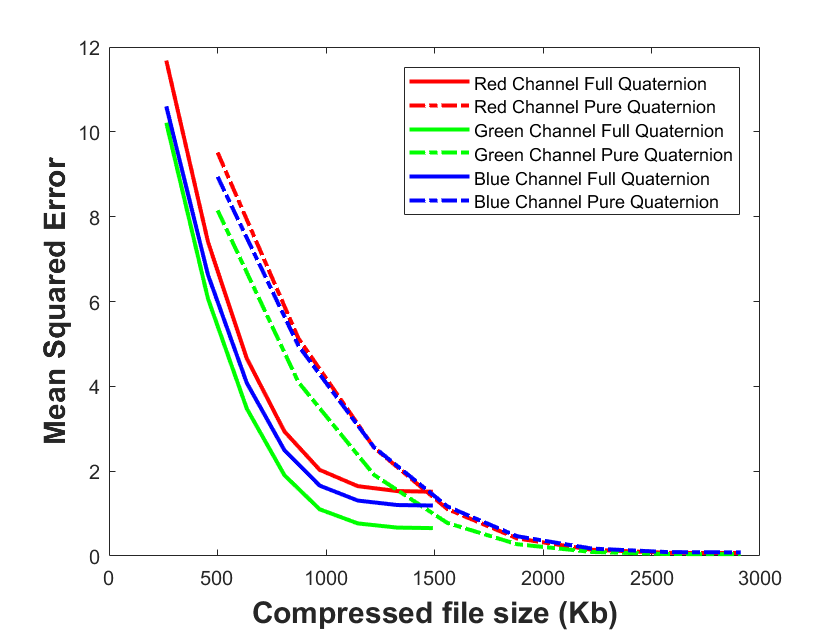}
  \caption{MSE of both of the compression methods for  $64 \times 64$ blocks}
  \label{fig:comparisonMSE64}
\end{figure}

MSE of color channels for $64 \times 64$ blocks are given in figure \ref{fig:comparisonMSE64}. The figure is clearly indicative of better performance of our proposed method concerning the error of reconstructed channels. The result of this test for $32 \times 32$ blocks is extremely similar to figure\ref{fig:comparisonMSE64}.

\par
Based on the charts above, Not only does full quaternion representation lead to image processing in a holistic way, but also the salient refinement compared to the pure quaternion representation is the improvement in time- and space-efficiency.
\section{Conclusion}
Using cross-correlation of color images has motivated researchers to use quaternions for color image representation. Pure quaternion representation brought advantages to color image processing, especially in terms of auto- and cross-correlation of color images. Nevertheless, it introduced additional costs due to the extra fourth dimenstion. By using an autoencoder neural network, we presented a lossy model that encodes a color image into full quaternion matrix. Apart from salient advantages of quaternions, all four components of quaternion matrix are correlated in our proposed approach and the computation costs reduce significantly due to the reduction in the number of matrix columns. The average PSNR of 41.2496 and SSIM of 0.9931 on UCID dataset indicate that there is only a slight amount of error in RGB image reconstruction, which is because of using a global model that tries to convert every pair of pixels into a quaternion number. Future research on this model can improve its accuracy and may contribute to a lossless model.
\par
Furthermore, we presented a QSVD-based color image compression method using our proposed full quaternion representation to put it  into practice as a case study. The comparison of this method with compression based on pure quaternion representation was completely on the favor of our proposed approach with refinements on time, space and image quality. The compression using full quaternion representation was almost two times faster, and the quality was higher in the same compressed file size. Further research on applications of full quaternion representation could be conducted in the fields of deep learning networks and video compression.
\section{Acknowledgment}
The authors thank Payman Moallem for the helpful discussion on the possible arrangement of pixels for training the autoencoder.



\end{document}